# Experimental study of a planar atmospheric-pressure plasma operating in the microplasma regime

A. J. Wagner,[1,2] D. Mariotti,[1,3,*] K. J. Yurchenko,[1] and T. K. Das[4]

[1]*Department of Microelectronic Engineering, Rochester Institute of Technology, Rochester, New York 14623, USA*
[2]*Department of Chemical Engineering and Materials Science, University of Minnesota, Minneapolis, Minnesota 55455, USA*
[3]*Nanotechnology & Advanced Materials Research Institute (NAMRI), University of Ulster, Newtownabbey BT37 0QB, United Kingdom*
[4]*Department of Mechanical Engineering, Rochester Institute of Technology, Rochester, New York 14623, USA*

Electrical characterization of a nonthermal radio-frequency atmospheric-pressure microplasma in a parallel plate configuration has shown that reducing electrode gap into the submillimeter range increases current and power density at a reduced voltage as compared to similar plasmas at larger electrode gaps which have no gap dependence. Calculation of sheath thickness and electric fields in the sheath and in the bulk demonstrate a dependence on the electrode gap as it is reduced into the submillimeter regime, indicating a distinct regime of operation.

Investigation of nonequilibrium atmospheric-pressure plasmas has grown over the last 15 years as their benefits are utilized in many applications [1–5]. In particular, atmospheric-pressure microplasmas (AMPs) broadly defined as plasmas confined to submillimeter cavities have attracted much attention. Although particular properties are recognized to be specific to AMPs, there is a lack of understanding of the conditions required to generate a "*microplasma regime*" (MPR) and the physical processes that determine their unique properties. However, these properties have inspired a widespread effort to study AMPs for many different applications including nanofabrication [1,6–12], propulsion [13,14], and photonics [15], among others [16–21]. In pursuit of understanding AMP properties, this work focuses on current-voltage (*I-V*) characteristics and sheath behavior derived from these characteristics of rf plasmas confined between electrodes at submillimeter gaps. The parallel-plate configuration represented an experimental milestone in developing the theories for low-pressure plasma and is an essential step to develop a basic scientific understanding of microplasma physics.

In large rf plasmas, the sheath thickness is small compared to the interelectrode spacing and the capacitive sheath can fully develop. As the distance between the electrodes is reduced to sizes comparable to the sheath thickness, changes in the plasma processes must occur for the plasma to be sustained. Results of simulations reported in the literature show that while sheath scaling with current density is preserved, sheath thickness in AMPs also scale with the electrode gap [22]. When this happens, the electron energy distribution function (EEDF) becomes heavily time modulated with electron trapping no longer possible in the same way as observed for larger gaps. Also gas ionization is affected as the sheath-bulk balance is disrupted with the creation of a structure very close to a sheath-only plasma. This phenomenon is strictly related to the MPR and it can be argued that in AMPs the $\gamma$ mode can be easily sustained versus the $\alpha$ mode. However, these results have not been confirmed experimentally and due to the dominant capacitive nature of AMP as shown in this Rapid Communication, a regime that differs from both $\alpha$ and $\gamma$ mode has to be considered a possibility.

Here we report on the experimental results of a study of the sheath properties as the electrode gap is reduced into the submillimeter range. We report on the *V-I* characteristics and related measurements showing evidence for the existence of a distinct regime, the MPR. Furthermore we provide evidence that in the MPR, the sheath maintains its primary function of coupling energy to the electrons to sustain the bulk plasma. We also confirm that in this regime, different plasma mechanisms are occurring which redefine the bulk-sheath boundary.

The plasma apparatus depicted in Fig. 1 consists of two planar aluminum electrodes. The powered electrode (50.8

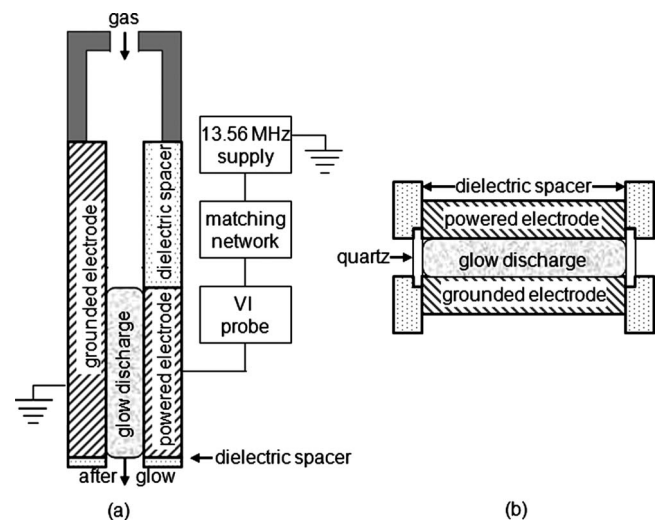

FIG. 1. Schematic of the experimental setup from the (a) side and (b) bottom. Two aluminum electrodes are supported on their sides by ceramic spacers with inlayed Corning 7980 quartz windows, enabling spectroscopic study along the length of the discharge.

---

*d.mariotti@ulster.ac.uk

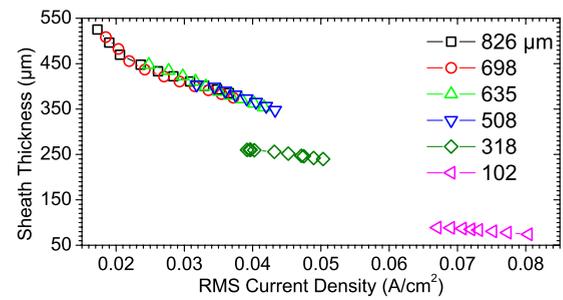

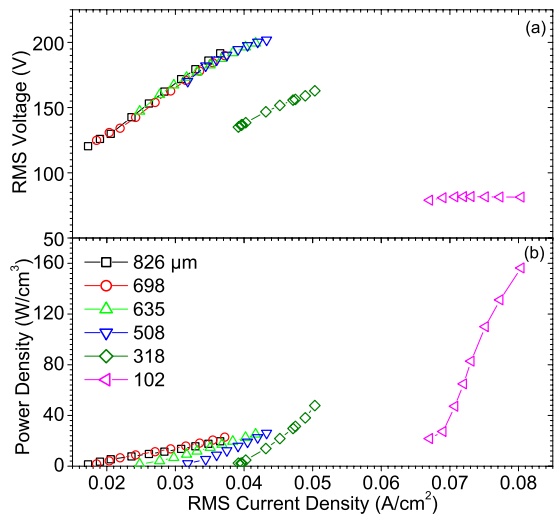

FIG. 2. (Color online) Experimental (a) voltage-current density and (b) power-current density characteristics versus current density of the discharge for different electrode gaps (power density is calculated considering the whole discharge volume).

mm wide and 101.6 mm long) is fixed, while a grounded electrode (50.8 mm wide and 152.4 mm long) is positioned at a variable distance (102–826 $\mu$m) from the powered electrode. The gap is sealed on each side by a ceramic plate and an inlaid Corning 7980 window extending beyond the length of the powered electrode to allow for spectral and spatial imaging. The powered electrode is connected to a Comdel CDX-2000 13.56 MHz rf power supply through an MKS MWH-5 impedance matching network. Helium gas (Airgas 99.999% purity) is introduced into a diffusion box prior to flowing through the gap between the electrodes.

The distance between the electrodes was established by inserting a spacer of the desired thickness while securing the grounded electrode in position. The spacer consisted of a thin polyethylene terephthalate glycol foil of the same area as the grounded electrode so that a uniform gap could be maintained. Although control of the electrode gap can be improved, repeated experiments showed consistent results with a very uniform plasma distribution over the full discharge area. The electrical properties of the discharge were studied using an MKS broadband $V/I$ probe rf impedance analyzer with the rf sensor connected in series with the matching network via coaxial wire 40 mm from the powered electrode. High-magnification optical photography, gas temperature measurements, and optical emission studies have been performed and will be presented in a future publication. However, it is important to mention here that based on these results it is confirmed that the discharge produced is a low-temperature nonequilibrium diffuse glow plasma.

In order to compare electrical characteristics and to carry out the analysis of the sheath thickness at different electrode gaps, a constant linear gas velocity of 20 m/s has been selected for all measurements. The $V$-$I$ and power-current ($P$-$I$) characteristics for six electrode gaps are reported in Figs. 2(a) and 2(b), respectively. The electrical measurements clearly show a common trend for electrode gap larger than 318 $\mu$m, whereby for a given voltage the same current den-sity is produced. However it is observed that as the gap is reduced below 508 $\mu$m the same relationship is not maintained and a much higher current is produced at the same electrode voltage. This different behavior is an indicator of the different plasma processes needed in order to maintain the discharge.

FIG. 3. (Color online) Calculated sheath thickness versus current density of the discharge for different electrode gaps.

The relation of the power density with current density [Fig. 2(b)] reveals an interesting feature of smaller gaps. The power density was calculated considering the whole discharge volume given by 50.8 mm $\times$ 101.6 mm $\times$ (electrode distance). The range of power density achievable when the electrode gap is smaller than 508 $\mu$m is much higher than the power observed at larger gaps, while also operating at a lower applied voltage. At 318 and 102 $\mu$m, the power density is an order of magnitude higher than that for a larger gap, indicating a much more efficient energy transfer to charged particles and possibly to inelastic collisions.

Analysis of the sheath thickness can aid in understanding the transition from a large-gap plasma into an MPR. The plasma sheath thickness has been estimated from the electrical measurements which have shown the presence of a capacitive component and a resistive one. The experimental capacitive component has been associated with the two sheath regions and the resistive component with the bulk plasma [23,24]. Here we will consider the total sheath capacitance given by the series of the two sheath capacitance and the total sheath thickness as the sum of the thickness of the two sheath regions. The total capacitance $C_T$ of the plasma and the time-averaged total sheath thickness $s$ are, respectively,

$$C_T = \frac{I}{\omega V}\sqrt{1+\tan(\phi)^{-2}} \quad \text{and} \quad s = \frac{\varepsilon_0 A}{C_T}, \quad (1)$$

where $V$ is the voltage across the gap, $\omega$ is the angular frequency of the applied power, $\phi$ is the measured $I$-$V$ phase angle in radians, $\varepsilon_0$ is the vacuum permittivity, and $A$ is the cross sectional area of the plasma, i.e., the area of the powered electrode.

Performing calculations for the measurements in Fig. 2(a), we obtain the total sheath thickness at different electrode gaps as shown in Fig. 3, confirming the existence of a transition between 508 $\mu$m gap and 318 $\mu$m gap. For large gaps, it is well accepted that the sheath thickness is independent of the plasma electrode gap and that the thickness only decreases with increasing current density. As the gap is reduced the bulk plasma is traditionally expected to shrink, leaving the sheath thickness unaltered [22,24]. This behavior has never been verified experimentally for submillimeter electrode gaps at atmospheric pressure and recent simulations performed by Shi and Kong have predicted that the sheath shrinks [22]. Our results confirm that the sheath becomes dependent on the electrode gap at some point between 508 and 318 $\mu$m, as seen in Fig. 3. It is evident that while the sheath is allowed to fully develop in a large gap, there exists a point at which a delicate balance between bulk plasma and sheath comes into play. This transition must correspond to changes in the contributions of the different plasma processes and a distinct regime of atmospheric plasma is entered. The transition from a large plasma to a microplasma is also apparent when comparing sheath thickness to the electrode gap at a constant current density (e.g., at 600 A/m$^2$), seen in Fig. 4. The sheath thickness remains constant above 318 $\mu$m electrode gap and decreases for gaps below 508 $\mu$m.

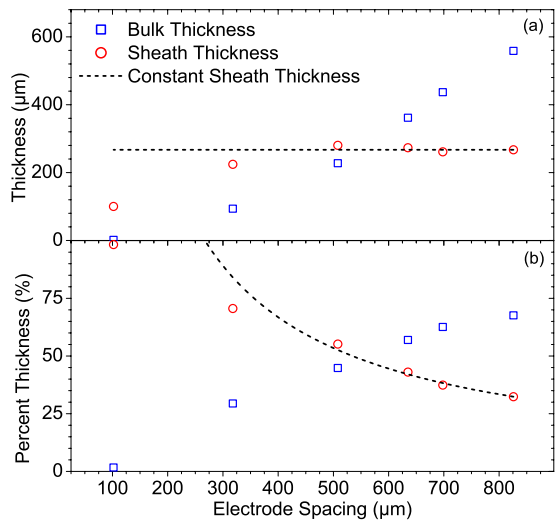

FIG. 4. (Color online) Calculated sheath and bulk thicknesses versus (a) gap size and (b) as a percent of the electrode spacing for a current density of 600 A/m$^2$. The dotted line represents constant sheath thickness.

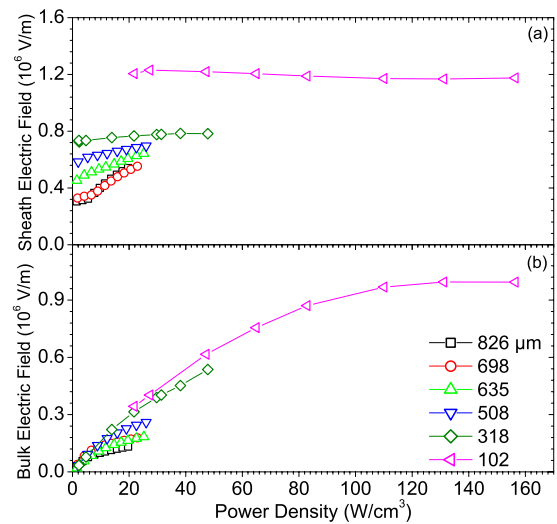

FIG. 5. (Color online) Calculated (a) sheath and (b) bulk electric field with respect to power density at different electrode gaps.

Further evidence is provided by observing the electric fields across the bulk plasma and across the sheath, seen in the top and bottom of Fig. 5, respectively. The electric field has been determined by calculating the voltage drop across the sheath and across the bulk from the complex impedance of the plasma divided by the calculated sheath and bulk thicknesses, respectively. The voltage amplitude has been considered for the calculations so that the electric field corresponds to the maximum electric field in an rf cycle.

Scaling from larger to smaller electrode spacings, the plasma experiences a change in state as the power density is increased drastically due to efficient inelastic electron collision [see below and Fig. 2(a)]. However it appears that an asymptotic limit for the sheath electric field is reached between 318 and 508 $\mu$m electrode gap which is likely related to the half-rf cycle during which the momentary positive electrode is negatively charged with electrons. On the contrary the bulk electric field, which is generally very weak in large-gap plasmas, seen in Fig. 5(b), becomes very high when the electrode spacing is reduced below 508 $\mu$m. In particular at 102 $\mu$m the electric field in the bulk is comparable to the sheath electric field.

The bulk plasma is generally defined as a time-averaged electrically neutral region of high plasma density that is insulated from the electrodes by the sheaths [24]. While the electric field in this region remains small at large gaps, the high electron density allows high rates of collisions, sustaining the plasma. As the electrode spacing is reduced, the bulk electric field increases and electron drift to the momentary anode increases. As a result, the electron density and bulk thickness decrease while the average electron energy increases. It follows that if at large gap ionization is sustained by high electron number density, at small gaps, the redistribution of the EEDF with a higher-energy electron group results in an increased efficiency in the sustainment of the plasma.

From our experimental results it is evident that as the electrode gap is reduced a transition into a different electrical behavior is observed by measurements. The regime that we have termed the MPR has specific characteristics: (a) the voltage required to sustain a given current density scales with the electrode gap; (b) the maximum achievable power-density scales inversely with the electrode gap; (c) the sheath thickness scales with the electrode gap and the majority of the electrode spacing is capacitive in nature; (d) the electric field in the bulk plasma is comparable to the electric field in the sheath and varies with the gap. These characteristics derived from the electrical measurements are sufficient to define the existence of a regime that differs from what is generally observed at larger electrode gaps. Although it was natural to believe that the plasma could enter the $\gamma$ mode as the electrode gap was reduced [22], we have now sufficient experimental evidence to show that this does not happen. Despite similarities, the MPR differs from the $\gamma$ mode in a few aspects and in particular it can be observed that the MPR is largely capacitive with no observation of instabilities or arcing.

Ionization in the MPR is strongly determined by energetic electrons given the very high electric field in both the sheath and the bulk plasma (Fig. 5). Taking current to be $j=qn_e\mu E$, where $q$ is the elementary charge, $n_e$ is the electron density, $\mu$ is the electron mobility, and $E$ is the electric field, it can be shown that the volume-averaged electron density will tend to decrease due to the increased electric field. It follows that electrons in the MPR are less in number but more efficient in the ionization process. Given the nonlinear nature of inelastic collisions it can be safely concluded that the efficiency of electron-driven reactions and in particular the formation of radicals is enhanced in the MPR and can possibly be optimized for specific reactions. This is also supported by a drastic increase in the power-density ranges reached in AMP [Fig. 2(b)].

The structure of the plasma in the MPR remains divided into three regions, the bulk and the two plasma sheaths; however, these regions have assumed different characteristics. The bulk plasma is the smaller part of the full electrode gap ($<50\%$) with a high electric field dependent on the gap. In this region the electron density is expected to be sufficiently high to largely determine the overall rate of inelastic processes. The sheath seems to have some unchanged properties if compared with large-gap plasmas with the exception of the sheath-bulk boundary. Given the numerical results in [22,25] with regards to the electron density and energy distributions, it is reasonable to assume that an energetic electron group exists that acts similarly to a particle beam oscillating between the two electrodes and within the bulk plasma. Here we speculate that at sufficiently small gap, the group of energetic electrons can be almost identified with the sheath-bulk boundary where electrons accumulate. This would also justify the electric field in the bulk being comparable to the electric field in the sheath (Fig. 5) and would lead to the conclusion that these microplasmas are overall electropositive.

In this Rapid Communication we have provided experimental evidence of a different plasma regime due to spatial confinement in one dimension. We reported on the electrical measurements and calculations of plasma properties that clearly show a marked transition into this distinct regime. The MPR is unique from previously observed plasma modes as discussed in our analysis. We expect that further studies will elucidate more details of the MPR, particularly with respect to a two-dimensional (2D) and three-dimensional (3D) confinement and to other plasma control parameters. These experimental results together with simulation results reported in [22,25] open the door to additional theoretical and experimental studies which will contribute to greater understanding of AMP. Moreover, we hope new technology-related advances will improve available diagnostic capabilities for this scientifically and technologically important type of plasma.

The authors would like to thank Rick Church, David Coumou, and Todd Hackelman (MKS Instruments, Rochester, NY, USA) and Bruce Tolleson of the Semiconductor and Microsystems Fabrication Laboratory of the Rochester Institute of Technology (Rochester, NY, USA) for their invaluable technical support. This work was partially supported by Applied Materials (Santa Clara, CA, USA), MKS Instruments (Rochester, NY, USA), and the NSF under Awards No. EECS-0530575 and No. EECS-0839961.